\documentstyle[preprint]{ptptex}
\newcommand{\beq}{\begin{equation}}
\newcommand{\eeq}{\end{equation}}
\newcommand{\beqa}{\begin{eqnarray}}
\newcommand{\eeqa}{\end{eqnarray}}

\title{%        %You can use \\ for explicit line-break
Vanishing Next-to-Leading Corrections to\\
the $\beta$-Function of the SUSY $CP^{N-1}$ Model\\
in Three Dimensions
}
\author{%       %Use \sc for the family name
Takeo {\sc Inami},
Yorinori {\sc Saito}
and
Masayoshi {\sc Yamamoto}%\footnote{E-mail address: yamamoto@phys.chuo-u.ac.jp} 
}

\inst{%         %Affiliation, neglected when [addenda] or [errata]
Department of Physics,
Faculty of Science and Engineering\\
Chuo University,
Tokyo 112-8551, Japan
}

\recdate{%      %Editorial Office will fill in this.
}

\abst{%       %this abstract is neglected when [addenda] or [errata]
We study the ultraviolet properties of the supersymmetric $CP^{N-1}$ sigma model 
in three dimensions to next-to-leading order in the $1/N$ expansion. 
We calculate the $\beta$-function to this order
and verify that it has no next-to-leading order corrections.
}

\begin{document}

\maketitle

\section{Introduction}

%\noindent {\it 1. Introduction}~~~ 
The supersymmetric $CP^{N-1}$ sigma model in low dimensions has many interesting 
features.
In two dimensions this model shares a few physical properties with 
supersymmetric gauge theories in four dimensions.\cite{rf:Witten,rf:DAdda}
Namely, perturbatively it is asymptotically free,
and non-perturbatively it has instantons.
Moreover, the one-loop exact $\beta$-function was found using instanton methods
in supersymmetric K\"ahler sigma models
which contain the supersymmetric $CP^{N-1}$ model as a special case.\cite{rf:Morozov}
In three dimensions, nonlinear sigma models are perturbatively non-renormalizable, 
but they are argued to be renormalizable in the $1/N$ expansion.
\cite{rf:Arefeva1,rf:Arefeva3,rf:Rosenstein}
The renormalization of the $N=1$ supersymmetric $O(N)$ sigma model
was worked out explicitly to next-to-leading order in $1/N$.\cite{rf:Koures}
Elucidating the possible similarity between the supersymmetric $CP^{N-1}$ sigma model
in three dimensions
and supersymmetric gauge theories in five dimensions
is also an interesting problem.

The aim of this letter is to study the ultraviolet (UV) properties of the $N=2$ 
supersymmetric $CP^{N-1}$ sigma model in three dimensions using the $1/N$ expansion.
Nonlinear sigma models in three dimensions are plagued by
a number of various power divergences in the cutoff $\Lambda$.
We investigate how such UV divergences may combine to cancel out in the model.
To this end we use the cutoff regularization.
In the $N=1$ supersymmetric $O(N)$ sigma model,
the $\beta$-function was found to be zero in dimensional regularization.\cite{rf:Koures}
We have confirmed that the fixed point of that model has next-to-leading order corrections
in the cutoff regularization.

A new method relying on conformal symmetry has been used to study nonlinear sigma models
beyond the leading order in $1/N$.\cite{rf:Gracey,rf:Ciuchini}
This method has been argued to be valid for nonlinear sigma models in general spacetime
dimensions $d<4$,
and it has been used to compute critical exponents in bosonic and
supersymmetric nonlinear sigma models in $d=3$.\cite{rf:Gracey,rf:Ciuchini}
We calculate the $\beta$-function of the supersymmetric $CP^{N-1}$ model
in the cutoff regularization.
We verify that the $\beta$-function has no next-to-leading order corrections.
Our results are compared with the previous results.\cite{rf:Koures,rf:Ciuchini}

\section{The Supersymmetric $CP^{N-1}$ Model in Three Dimensions}

%\noindent {\it 2. The Supersymmetric $CP^{N-1}$ Model in Three Dimensions}~~~ 
We begin by outlining the $N=2$ supersymmetric $CP^{N-1}$ sigma model 
in three dimensions.
We use the $N=1$ complex scalar superfields
$\Phi_j=n_j+\bar{\theta}\psi_j+(1/2)\bar{\theta}\theta F_j$,
where $j=1,\cdots,N$.
The action of the model is written in the supergauge-invariant
form \cite{rf:DAdda,rf:Ciuchini}
\beq
S=\int d^3xd^2\theta ~\overline{\nabla\Phi_j}\nabla\Phi_j,   
\label{2.2}
\eeq
with the constraint
$\bar{\Phi}_j\Phi_j=N/2g$.
Here $g$ is the coupling constant and $\nabla_\alpha$ is the gauge covariant
supercovariant derivative
$\nabla_\alpha=D_\alpha-iA_\alpha$,
where $A_\alpha$ is a spinor superfield and $D_\alpha$ is the supercovariant
derivative
$D_\alpha=\partial/\partial \bar{\theta}_\alpha
-i(\gamma^\mu \theta)_\alpha\partial_\mu$.
Our convention for the gamma matrices is given by $\gamma^0=\sigma^2$, $\gamma^1=i\sigma^3$
and $\gamma^2=i\sigma^1$. 

We use the Wess-Zumino gauge 
in which $A_\alpha$ has the form
$A_\alpha=i(\gamma^\mu\theta)_\alpha A_\mu
+(1/2)\bar{\theta}\theta \omega_\alpha$,
where $A_\mu$ is a $U(1)$ gauge field and $\omega_\alpha$ is a Majorana spinor.
We introduce a real scalar superfield
$\Sigma=\sigma+\bar{\theta}\xi+(1/2)\bar{\theta}\theta \alpha$
as the Lagrange multiplier.
The action can then be written as
\beq
S=\int d^3xd^2\theta \left[\overline{\nabla\Phi_j}\nabla\Phi_j
+2\Sigma\left(\bar{\Phi}_j\Phi_j-\frac{N}{2g}\right)\right].
\label{2.8}
\eeq
In component fields, the Euclidean action is given after eliminating $F_j$ by
\beqa
S&=&\int d^3x \biggl(-\bar{n}_j{\partial}^2n_j+i\bar{\psi}_j\partial\!\!\!/\psi_j
+\frac{N}{2g}\alpha+\sigma^2\bar{n}_jn_j-\alpha\bar{n}_jn_j+\sigma\bar{\psi}_j\psi_j
\nonumber \\
&&-iA_\mu\bar{n}_j\stackrel{\leftrightarrow}{\partial}_\mu n_j
-A_\mu\bar{\psi}_j\gamma_\mu\psi_j+A_\mu A_\mu\bar{n}_jn_j+\bar{n}_j\bar{c}\psi_j
+n_j\bar{\psi}_jc\biggr),
\label{2.9}
\eeqa
where $c=\xi+i\omega/2$ is a complex fermion.
This model is known to have $N=2$ supersymmetry.
\cite{rf:Ciuchini,rf:Aoyama}

\section{The Leading Order}

%\noindent {\it 3. The Leading Order}~~~ 
The generating functional of the model in the Euclidean notation is
\beqa
Z(J_j,\bar{J}_j,\eta_j,\bar{\eta}_j)
&=&\int {\cal D}n_j{\cal D}\bar{n}_j{\cal D}\psi_j{\cal D}\bar{\psi}_j
{\cal D}\alpha {\cal D}\sigma {\cal D}A_\mu {\cal D}c{\cal D}\bar{c}
\nonumber \\
&&
\times \exp\left[-S+\int d^3x(\bar{J}_jn_j+\bar{n}_jJ_j
+\bar{\eta}_j\psi_j+\bar{\psi}_j\eta_j)\right].
\label{2.10}
\eeqa
Performing the integrations over the fields $\psi_j$, $\bar{\psi}_j$, $n_j$ and 
$\bar{n}_j$ 
we obtain
\beq
Z(J_j,\bar{J}_j,\eta_j,\bar{\eta}_j)
=\int {\cal D}\alpha {\cal D}\sigma {\cal D}A_\mu {\cal D}c{\cal D}\bar{c}~\exp(-S_{\rm eff}).
\label{2.11}
\eeq
The effective action $S_{\rm eff}$ is given by
\beqa
S_{\rm eff}&=&N{\rm Tr}\ln (\Delta_B-\bar{c}\Delta_F^{-1}c)-N{\rm Tr}\ln\Delta_F
-\int d^3x\biggl[(\bar{J}_j-\bar{\eta}_j\Delta_F^{-1}c)
\nonumber \\
&&
\times (\Delta_B-\bar{c}\Delta_F^{-1}c)^{-1}(J_j-\bar{c}\Delta_F^{-1}\eta_j)
+\bar{\eta}_j\Delta_F^{-1}\eta_j-\frac{N}{2g}\alpha\biggr],
\label{2.12}
\eeqa
where
$\Delta_F=i\partial\!\!\!/-\gamma_\mu A_\mu+\sigma$
and $\Delta_B=-\partial^2-iA_\mu\stackrel{\leftrightarrow}{\partial}_\mu+A_\mu A_\mu
+\sigma^2-\alpha$.
Performing the Legendre transformation and setting all fields to constants,
we obtain the effective potential
\beqa
V&=&N\biggl[\bar{v}v(\langle\sigma\rangle^2-\langle\alpha\rangle)+\frac{1}{2g}\langle\alpha\rangle
\nonumber \\
&&
+\int\frac{d^3k}{(2\pi)^3}\left(\ln (k^2+\langle\sigma\rangle^2-\langle\alpha\rangle)
-{\rm tr}\ln (-k\!\!\!/+\langle\sigma\rangle)\right)\biggr],
\label{2.15}
\eeqa
where $v=\langle n_N\rangle/\sqrt{N}$.
The fields which are not in (\ref{2.15}) have been set to zero.

The vacuum of the model is fixed by the stationary conditions
\beqa
&& \bar{v}\langle\sigma\rangle^2=v\langle\sigma\rangle^2=0,
\label{2.16}\\
&& \bar{v}v-\frac{1}{2g}+\int\frac{d^3k}{(2\pi)^3}\frac{1}{k^2+\langle\sigma\rangle^2}=0.
\label{2.17}
\eeqa
We look for the supersymmetric vacuum and have set $\langle\alpha\rangle=0$.
The UV divergences present in the integral in (\ref{2.17}) can be dealt 
with by renormalization.
Introducing a scale parameter $\mu$ and a renormalized coupling constant $g_R$,
the equation (\ref{2.17}) becomes
\beq
\bar{v}v-\frac{1}{2g_R}+\frac{\mu}{4\pi}-\frac{\langle\sigma\rangle}{4\pi}=0,
\label{2.18}
\eeq
where $\langle\sigma\rangle$ is positive.
The renormalized and bare coupling constants are related by
\beq
\frac{1}{2g}=\frac{1}{2g_R}+\frac{\Lambda}{2\pi^2}-\frac{\mu}{4\pi},
\label{2.19}
\eeq
where $\Lambda$ is the momentum cutoff.
In terms of the dimensionless coupling constant defined by $\tilde{g}=g_R\mu$,
the $\beta$-function is
\beq
\beta(\tilde{g})=\tilde{g}\left(1-\frac{\tilde{g}}{2\pi}\right).
\label{2.20}
\eeq
This result is the same as that of the bosonic $CP^{N-1}$ model.\cite{rf:Cant}
The equations (\ref{2.16}) and (\ref{2.18}) imply the existence of two phases:
(i) for $\tilde{g}>2\pi$, $m\equiv \langle\sigma\rangle=2\pi\mu(1/2\pi-1/\tilde{g})$, 
$v=\bar{v}=0$ ($SU(N)$ symmetric phase); 
(ii) for $\tilde{g}<2\pi$, $\langle\sigma\rangle=0$,
$\bar{v}v=(\mu/2)(1/\tilde{g}-1/2\pi)$ ($SU(N)$ broken phase).
The ultraviolet properties of the model should be the same in two cases,
so we consider only the symmetric phase.

The model contains four kinds of auxiliary fields:
two scalars $\alpha$, $\sigma$, a $U(1)$ vector $A_\mu$, and a complex fermion $c$.
They all begin to propagate after taking into account
the quantum effects of $n_j$ and $\psi_j$ loops.
Considering the fact that $\sigma$ acquires a non-zero vacuum expectation value $m$,
we perform the shift
\beq
\sigma\to m+\sigma.
\label{sshift}
\eeq
The fields $\alpha$ and $\sigma$ mix as they propagate.
It is convenient to diagonalize their propagators by rewriting $\alpha$ as
\beq
\alpha\to\alpha+2m\sigma.
\label{ashift}
\eeq
The effective propagators of $\alpha$, $\sigma$, $A_\mu$ and $c$ 
can be obtained from the effective action (\ref{2.12})
after redefining the fields $\sigma$ and $\alpha$ as (\ref{sshift}) and (\ref{ashift}).  
They are given by
\beqa
&& D^\alpha(p)=-\frac{4\pi}{N}I(p),
~~D^\sigma(p)=\frac{4\pi}{N}\frac{1}{p^2+4m^2}I(p),
~~D^c(p)=\frac{8\pi}{N}\frac{p\!\!\!/-2m}{p^2+4m^2}I(p),
\nonumber \\
&& D^A_{\mu\nu}(p)=\frac{4\pi}{N}\frac{p^2\delta_{\mu\nu}-p_\mu p_\nu
-2m\epsilon_{\mu\nu\rho}p_\rho}{p^2(p^2+4m^2)}I(p),
\label{2.24}
\eeqa
where
$I(p)=\sqrt{p^2}/\arctan (\sqrt{p^2}/2m)$.
We have used the Landau gauge in deriving $D^A_{\mu\nu}(p)$.
We note that the mixing terms between $A_\mu$ and $\alpha$, $\sigma$ vanish
in the effective action (\ref{2.12}),
and such mixing terms arise in the two-dimesional model.\cite{rf:DAdda}

We have a few comments on the effective propagators.
$D^\alpha(p)$ has a branch cut but no poles.
$D^\sigma(p)$, $D^A_{\mu\nu}(p)$ and $D^c(p)$ have poles at $p^2=-4m^2$,
which correspond to bound states.
The term which involves $\epsilon_{\mu\nu\rho}$ in $[D^A_{\mu\nu}(p)]^{-1}$
is induced by the fermion loop.
In the $p\to 0$ limit, $D^A_{\mu\nu}(p)$ has the same form as the gauge field propagator
of the Maxwell-Chern-Simons theory, where the gauge field is massive.
In the present model, the gauge field mass is $2m$.

\section{Next-to-Leading Order Corrections}

%\noindent {\it 4. Next-to-Leading Order Corrections}~~~
Performing the shift (\ref{sshift}) in the action (\ref{2.9}), 
we obtain the Lagrangian
\beqa
{\cal L}&=&\bar{n}_j(-{\partial}^2+m^2)n_j+\bar{\psi}_j(i\partial\!\!\!/+m)\psi_j
+\frac{N}{2g}\alpha+\frac{Nm}{g}\sigma
\nonumber \\
&& -\alpha\bar{n}_jn_j+\sigma\bar{\psi}_j\psi_j+\sigma^2\bar{n}_jn_j
-iA_\mu\bar{n}_j\stackrel{\leftrightarrow}{\partial}_\mu n_j
\nonumber \\
&& -A_\mu\bar{\psi}_j\gamma_\mu\psi_j+A_\mu A_\mu\bar{n}_jn_j+\bar{n}_j\bar{c}\psi_j
+n_j\bar{\psi}_jc
+{\cal L}_{\rm CT}.
\label{3.1}
\eeqa
The fields, mass and coupling constant appearing in (\ref{3.1}) are renormalized quantities.
The transformation (\ref{ashift}) has been performed in renormalized quantities.
${\cal L}_{\rm CT}$ consists of the counterterms which are designed to eliminate all 
UV divergences due to loop effects.
${\cal L}_{\rm CT}$ is given explicitly by
\beqa
{\cal L}_{\rm CT}&=&\bar{n}_j(-C_1{\partial}^2+C_2m^2)n_j
+\bar{\psi}_j(iC_3\partial\!\!\!/+C_4m)\psi_j
+C_5\frac{N}{2g}\alpha+C_6\frac{Nm}{g}\sigma
\nonumber \\
&& -C_7\alpha\bar{n}_jn_j+2C_8m\sigma\bar{n}_jn_j
+C_9\sigma\bar{\psi}_j\psi_j+C_{10}\sigma^2\bar{n}_jn_j
-iC_{11}A_\mu\bar{n}_j\stackrel{\leftrightarrow}{\partial}_\mu n_j
\nonumber \\
&& -C_{12}A_\mu\bar{\psi}_j\gamma_\mu\psi_j+C_{13}A_\mu A_\mu\bar{n}_jn_j
+C_{14}\bar{n}_j\bar{c}\psi_j+C_{15}n_j\bar{\psi}_jc.
\label{ct}
\eeqa
We define the renormalization constants of the fields, mass and coupling constant by
\beqa
&& n_{0j}=Z_n^{1/2}n_j,~~~\psi_{0j}=Z_\psi^{1/2}\psi_j,
~~~\alpha_0=Z_\alpha\alpha,~~~\sigma_0=Z_\sigma\sigma,
\nonumber \\
&& A_{0\mu}=Z_AA_\mu,~~~c_0=Z_cc,
~~~m_0=Z_mm,~~~g_0=Z_gg,
\label{zfactor}
\eeqa
where the suffix $0$ denotes a bare quantity.
%The quantities $g_0$ and $g$ correspond to $g$ and $g_R$ in 3.

Renormalizability of the model can be assured by showing that no terms
which are not contained in the bare Lagrangian
appear in ${\cal L}_{\rm CT}$.
In addition,
the $C_i$ in (\ref{ct}) are related to the $Z$-factors introduced in (\ref{zfactor}).
%We are now studying this problem in detail.
Before completing this analysis, we calculate $Z_g$ by using the relations
\beq
1+C_1=Z_n,~~~1+C_5=Z_\alpha Z_g^{-1},~~~1+C_7=Z_\alpha Z_n.
\label{relation}
\eeq
The $C_i$ and $Z$-factors are expanded in $1/N$ as
$C_i=C_i^{(0)}+C_i^{(1)}+\cdots$,
$Z=Z^{(0)}+Z^{(1)}+\cdots$.
At leading order, the $Z$-factors are
$Z_n^{(0)}=Z_\psi^{(0)}=Z_\alpha^{(0)}=Z_\sigma^{(0)}=Z_A^{(0)}=Z_c^{(0)}=Z_m^{(0)}=1$
and
\beq
(Z_g^{-1})^{(0)}=\left(\frac{\Lambda}{\pi^2}-\frac{m}{2\pi}\right)g,
\label{zg0}
\eeq
which is derived from (\ref{2.19}) and (\ref{2.18}) in the symmetric phase
($v=0$, $\langle\sigma\rangle=m$).

Now we consider the next-to-leading order.
We have calculated the next-to-leading order corrections to the self-energies of $n_j$ and $\psi_j$ 
and those to the three-point vertex functions $\alpha\bar{n}_jn_j$ and $\sigma\bar{\psi}_j\psi_j$.
The loop graphs contributing to the self-energy of $n_j$ contain UV power divergences.
These divergences cancel out in the sum of all graphs.
The same result holds for the other vertex functions.
The remaining logarithmic divergences are removed by renormalization. 
The $Z$-factors are
\beqa
&&Z_n=1+\frac{4}{N\pi^2}\ln\frac{\Lambda}{\mu},
~~~Z_\psi=1-\frac{4}{N\pi^2}\ln\frac{\Lambda}{\mu},
\label{znzp}
\\
&&Z_\alpha=Z_\sigma=Z_m=1.
\label{zazszm}
\eeqa

Component fields of a single superfield should be dealt with
by a single renormalization constant in a manifestly supersymmetric scheme.
In our computation $Z_n$ and $Z_\psi$ have turned out to be unequal,
even though $n_j$ and $\psi_j$ are in the same superfield.
This is probably because $Z_n$ and $Z_\psi$ are gauge dependent and we use
the Wess-Zumino gauge.\cite{rf:Ciuchini}
The result that the auxiliary fields $\alpha$, $\sigma$ and the mass are not renormalized
is same as that in the case of the supersymmetric $O(N)$ sigma model.\cite{rf:Koures} 

\begin{figure}[t]
\begin{center}
\input{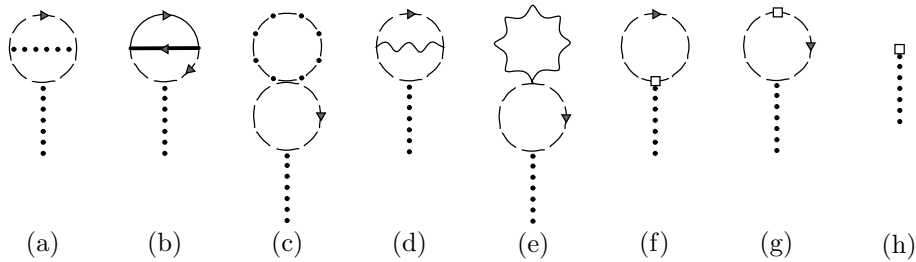} 
\end{center}
\bigskip
\caption{Next-to-leading order corrections to the $\alpha$-tadpole.
The dashed, solid, dotted, dash-dotted, wavy and thick lines denote
the propagators of $n_j$, $\psi_j$, $\alpha$, $\sigma$, $A_\mu$ and $c$, respectively.
The squares denote the counterterms.}
\label{tadpole}
\end{figure}

The $\beta$-function at next-to-leading order can be obtained by computing
the graphs of the $\alpha$-tadpole shown in Fig. \ref{tadpole}.
Remarkably, we have found that the sum of the graphs (a) to (e) of Fig. \ref{tadpole} is zero.
The sum of the graphs (f) and (g) is
\beq
(C_7^{(1)}-C_1^{(1)})N\int\frac{d^3k}{(2\pi)^3}\frac{1}{k^2+m^2}
=Z_\alpha^{(1)}N\left(\frac{\Lambda}{2\pi^2}-\frac{m}{4\pi}\right),
\label{correction}
\eeq
where we have used (\ref{relation}).
From (\ref{relation}) and (\ref{zg0}) the counterterm (h) is given by
\beqa
-C_5^{(1)}\frac{N}{2g}
&=&-\left(Z_\alpha^{(1)}(Z_g^{-1})^{(0)}+(Z_g^{-1})^{(1)}\right)\frac{N}{2g}
\nonumber \\
&=&-Z_\alpha^{(1)}N\left(\frac{\Lambda}{2\pi^2}-\frac{m}{4\pi}\right)-\frac{N}{2g}(Z_g^{-1})^{(1)}.
\label{counterterm}
\eeqa
Adding (\ref{correction}) and (\ref{counterterm}) the divergences cancel,
so we do not need $(Z_g^{-1})^{(1)}$ and obtain
\beq
(Z_g^{-1})^{(1)}=0.
\label{zg1}
\eeq
This implies that the $\beta$-function (\ref{2.20}) receives no corrections
at next-to-leading order.
We have checked that the same result can be derived by calculating corrections
to the $\sigma$-tadpole.
The present result is consistent with the result that the slope $\beta '(g_c)$
has no next-to-leading order corrections.\cite{rf:Ciuchini}

Our result for the $N=2$ supersymmetric $CP^{N-1}$ sigma model is in clear contrast to
that for the $N=1$ supersymmetric $O(N)$ sigma model.
We have found that in the $O(N)$ case there remains a linear divergence
in the next-to-leading order graphs of the tadpole
in the cutoff regularization.
We obtain $\beta (\tilde{g})=\tilde{g}[1-(\tilde{g}/4\pi)(1-4/N)]$.
This is different from the result $Z_g=1$ obtained
in dimensional regularization.\cite{rf:Koures}

In this letter we have derived the next-to-leading order corrections to the $\beta$-function,
but we have not completed the calculation of all divergent graphs at this order.
In the supersymmetric $O(N)$ sigma model,
all divergent graphs were calculated to next-to-leading order
and renormalizability was proved.\cite{rf:Koures}
We have verified that the four-point vertex function $(\bar{n}_jn_j)^2$ is finite
at next-to-leading order,
which is in accord with the renormalizability of the model.

\section{Discussion}

%\noindent {\it 5. Discussion}~~~
It is an important question
whether the absence of non-leading corrections to
the $\beta$-function persists to all orders in $1/N$,
namely whether the $\beta$-function of the model is leading order exact.
However, it is difficult to go beyond the next-to-leading order in the present approach,
and we need some new means to handle the problem, e.g. algebraically.

Comparing the present result with those in the bosonic $CP^{N-1}$ \cite{rf:Cant}
and $N=1$ supersymmetric $O(N)$ sigma model,
one may speculate that $N=2$ supersymmetry is responsible for the vanishing
of the next-to-leading order corrections to the $\beta$-function of the model.

\section*{Acknowledgements}
%\bigskip
We would like to thank S.~Yahikozawa for useful discussions
and M.~Sakamoto for a careful reading of the manuscript
and valuable comments.
We also wish to thank J. A. Gracey for a communication regarding
the supersymmetric $O(N)$ nonlinear sigma model.


\begin{thebibliography}{99}

\bibitem{rf:Witten} E.~Witten,
Nucl.~Phys. {\bf B149} (1979), 285.

\bibitem{rf:DAdda} A.~D'Adda, P.~Di Vecchia and M.~L\"uscher,
Nucl.~Phys. {\bf B152} (1979), 125.
 
\bibitem{rf:Morozov} A.~Y.~Morozov, A.~M.~Perelomov and M.~A.~Shifman,
Nucl.~Phys. {\bf B248} (1984), 279.

\bibitem{rf:Arefeva1} I.~Ya.~Aref'eva,
Theor.~Math.~Phys. {\bf 36} (1978), 573;
Ann.~Phys. {\bf 117} (1979), 393.

\bibitem{rf:Arefeva3} I.~Ya.~Aref'eva and S.~I.~Azakov, 
Nucl.~Phys. {\bf B162} (1980), 298.

\bibitem{rf:Rosenstein} B.~Rosenstein, B.~J.~Warr, and S.~H.~Park,
Nucl.~Phys. {\bf B336} (1990), 435.

\bibitem{rf:Koures} V.~G.~Koures and K.~T.~Mahanthappa,
Phys.~Rev. {\bf D43} (1991), 3428.

\bibitem{rf:Gracey} J.~A.~Gracey,
Nucl.~Phys. {\bf B352} (1991), 183.

\bibitem{rf:Ciuchini} M.~Ciuchini and J.~A.~Gracey, 
Nucl.~Phys. {\bf B454} (1995), 103.

\bibitem{rf:Aoyama} S.~Aoyama, 
Nucl.~Phys. {\bf B168} (1980), 354.

\bibitem{rf:Cant} R.~J.~Cant and A.~C.~Davis, 
Z.~Phys. {\bf C5} (1980), 299.

\end{thebibliography}
\end{document}